\documentclass{mn2e}
\input{epsf}

\title[Tidal tails of dwarf galaxies]
{Tidal tails of dwarf galaxies on different orbits around the Milky Way}

\author[E. L. {\L}okas et al.]
    {Ewa L. {\L}okas,$^{1}$ Grzegorz Gajda$^{2}$ and Stelios Kazantzidis$^{3}$
    \\
    \\
    $^1$Nicolaus Copernicus Astronomical Center, Bartycka 18, 00-716 Warsaw, Poland  \\
    $^2$Astronomical Observatory of the Jagiellonian University, Orla 171, 30-244 Cracow, Poland \\
    $^3$Center for Cosmology and Astro-Particle Physics; and Department of Physics; and Department of Astronomy,
    The Ohio State\\ University, Columbus, OH 43210, USA}

\begin{document}

\maketitle

\begin{abstract}
We present a phenomenological description of the properties of tidal tails forming around dwarf galaxies orbiting
the Milky Way. For this purpose we use collisionless $N$-body simulations of dwarfs initially composed of a disk
embedded in an NFW dark matter halo. The dwarfs are placed on seven orbits around the Milky Way-like host, differing
in size and eccentricity, and their evolution is followed for 10 Gyr. In addition to the well-studied morphological
and dynamical transformation of the dwarf's main body, the tidal stripping causes them to lose a substantial fraction
of mass both in dark matter and stars which form pronounced tidal tails. We focus on the properties of the stellar
component of the tidal tails thus formed. We first discuss the break radii in the stellar density profile
defining the transition to tidal tails as the radii where the profile becomes shallower and relate them to the classically
defined tidal radii. We then calculate the relative density and velocity of the tails at a few break radii as a function
of the orbital phase. Next, we measure the orientation of the tails with respect to an observer placed at the centre of
the Milky Way. The tails are perpendicular to this line of sight only for a short period of time near the pericentre.
For most of the time the angles between the tails and this line of sight are low, with orbit-averaged medians below 42
degrees for all, even the almost circular orbit. The median angle is typically lower while the maximum
relative density higher for more eccentric orbits. The combined effects of relative density and orientation
of the tails suggest that they should be easiest to detect for dwarf galaxies soon after their pericentre passage.
\end{abstract}


\section{Introduction}

Tidal tails of dwarf galaxies in the Local Group can provide important insight into the formation scenarios of
the Milky Way, M31 and their satellites. According to the presently accepted theories of structure formation
in the Universe, all galaxies formed by mergers and accretion of smaller objects. Tidal tails thus should be
omnipresent in the vicinity of the Milky Way as witnesses of the past accretion events. As most of the dwarf
galaxies were probably accreted at redshifts $z=1-2$ they should have had enough time to produce pronounced
tidal extensions as a result of their interaction with the Milky Way. It is also believed that the stellar
halo of our Galaxy formed from stars lost during such processes (Johnston 1998; Bullock \& Johnston 2005).
The features of tidal tails depend on the properties of the progenitor as well as the host and are thus
useful to constrain the Milky Way potential (Johnston et al. 1999a; Pe\~narrubia et al. 2006; Law \& Majewski 2010;
Koposov, Rix \& Hogg 2010; Ibata et al. 2013).

Most of the studies of the formation of tidal tails in the Local Group
to date were done in the context of tidal stripping of
Galactic globular clusters (Oh, Lin \& Aarseth 1992; Oh \& Lin 1992; K\"upper, Lane \& Heggie 2012).
Although smaller than dwarf galaxies in size, they are typically located much closer
to the Galactic centre than the dwarfs and therefore easier to disrupt. And indeed compelling evidence for
the presence of tidal tails around globular clusters has been found with the most prominent example being
the one of Palomar 5 (Odenkirchen et al. 2003).

\begin{table*}
\begin{center}
\caption{Orbital parameters of the simulated dwarfs. }
\begin{tabular}{ccccccccl}
Orbit & $r_{\rm apo}$ & $r_{\rm peri}$ & $r_{\rm apo}/r_{\rm peri}$ & $T_{\rm orb}$ & $t_{\rm la}$
& $n_{\rm peri}$ & $R_{\rm tail}$ & Colour \\
      & (kpc)         & (kpc)          &                            & (Gyr)         & (Gyr)
&                &    (kpc) &   \\
\hline
O1  &   125 &    25   &  \ \   5        & 2.09 & \ \  8.35  & 5 & $\,   12.7 $    &     green   \\
O2  &\ \ 85 &    17   &  \ \   5        & 1.28 & \ \  8.95  & 8 & $\,   10.7 $    &     red     \\
O3  &   250 &    50   &  \ \   5        & 5.40 & \ \  5.40  & 2 & $\,   34.2 $    &     blue    \\
O4  &   125 &\ \ 12.5 &  $\,  10$       & 1.81 & \ \  9.05  & 6 & $\,   11.0 $    &     orange  \\
O5  &   125 &    50   &\ \ \ \ 2.5      & 2.50 & $\, 10.00$ & 4 & $\,   11.7 $    &     purple  \\
O6  &\ \ 80 &    50   &\ \ \ \ 1.6      & 1.70 & \ \  8.50  & 6 & \ \    7.6      &     brown   \\
O7  &   250 &\ \ 12.5 &  $\,  20$       & 4.55 & \ \  9.10  & 2 & $\,   21.9 $    &     black   \\
\hline
\label{parameters}
\end{tabular}
\end{center}
\end{table*}

The main reason why tidal tails of dwarf galaxies were not as yet extensively studied is probably the lack of
good photometric and kinematic measurements. Although some indication of the presence of the tails was found
in Ursa Minor (Mart{\'\i}nez-Delgado et al. 2001), Carina (Mu\~noz et al. 2006) or Leo I (Sohn et al. 2007)
and hints of stream-like overdensities have also been detected around some of the ultra-faint
satellites of the Milky Way (Sand et al. 2012), the presence of tidal tails around most Local Group dwarf
galaxies is still under debate ({\L}okas et al 2012; Frinchaboy et al. 2012).
A notable counter example is the Sagittarius dwarf whose tails were detected beyond any doubt (Majewski et al. 2003)
and studied in detail by many authors (e.g. Johnston, Spergel \& Hernquist 1995; Johnston et al. 1999b;
Helmi \& White 2001; Helmi 2004; Law, Johnston \& Majewski 2005; Law \& Majewski 2010; {\L}okas et al. 2010).
Interestingly, streams of stars with probable tidal origin but without obvious progenitor identified have been
discovered in the Local Group, e.g. the Orphan stream (Belokurov et al. 2007) or the Monoceros stream
(Newberg et al. 2002). There is also compelling evidence of tidal features around dwarf galaxies and streams beyond our
immediate cosmic neighbourhood (e.g. Mart{\'\i}nez-Delgado et al. 2010, 2012; Koch et al. 2012)

Klimentowski et al. (2009a) used $N$-body simulations to study the properties of tidal tails forming
around a dwarf galaxy interacting with a Milky Way-like host. The dwarf was placed on an eccentric orbit typical
of Milky Way satellites with apo- to pericentre distances $r_{\rm apo}/r_{\rm peri}=120/25$ kpc. Contrary to earlier
studies which assumed a spherical, single-component progenitor (e.g. Johnston, Choi \& Guhathakurta 2002;
Choi, Weinberg \& Katz 2007), the dwarf galaxy
was initially composed of a stellar disk embedded in a dark matter halo and was thus akin to dwarf irregular galaxies
believed to be progenitors of the present-day dwarf spheroidal (dSph) galaxies according to the tidal stirring scenario
(Mayer et al. 2001). Their study was the first to address the issue of the formation of tidal tails in the
context of this scenario and without the assumption of the tidally stripped dwarf being spherical.

The most interesting
finding of this work was the conclusion related to the orientation of the tails with respect to an observer placed
near the centre of the Milky Way. It turned out that the tails are typically inclined by a rather small angle to
this line of sight and therefore may not be easy to detect except for a brief period of time when the dwarf galaxy
is near the pericentre of its orbit. This finding also had important implications for the mass modelling of
dSph galaxies since tidal tails oriented along the line of sight tend to significantly contaminate kinematic
samples used to determine masses (Klimentowski et al. 2007).

Although typical for subhaloes accreted at redshift $z=1-2$, the orbit considered by Klimentowski et al. (2009a)
was only one of the whole spectrum of orbits found in simulations of the Local Group (Diemand, Kuhlen \& Madau 2007;
Klimentowski et al. 2010). Here we extend the work of Klimentowski et al. (2009a) to include orbits of different
size and eccentricity while still working within the context of the tidal stirring model, i.e. assuming realistic
progenitor disky dwarfs. Our initial conditions also differ slightly in the way we model the dwarf and the host
galaxy.

The paper is organized as follows. In section 2 we shortly describe the simulations used in this work. Section 3 is
devoted to the description of the stellar density profiles of the simulated dwarfs and in their immediate vicinity
with the purpose of determining the radius of transition between the dwarf's main body and the tidal tails. In section
4 we discuss the density and kinematics of the tails as a function of their orbital phase and in section 5 we look at
the orientation of the tails with respect to the line of sight of an observer placed near the centre of the Milky Way.
The discussion follows in section 6.

\section{The simulations}

The collisionless $N$-body
simulations we use here were presented in detail in Kazantzidis et al. (2011) and {\L}okas et al. (2011). These
works focused on the study of the formation of dSph
galaxies via tidal stirring of
rotationally supported dwarfs in the vicinity of Milky Way-sized hosts.

\begin{figure}
\begin{center}
    \leavevmode
    \epsfxsize=7cm
    \epsfbox[10 10 200 202]{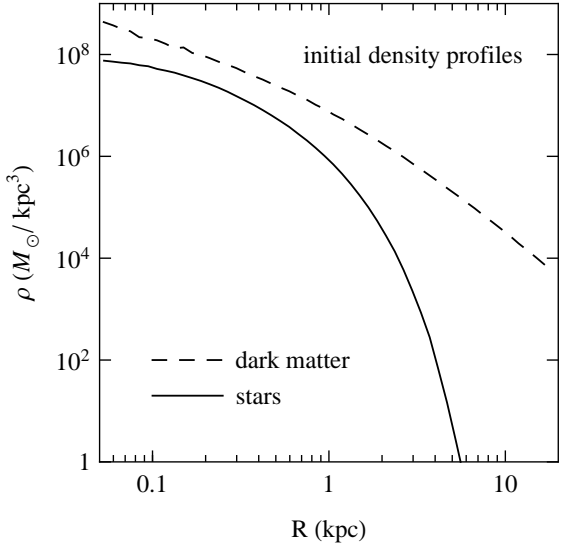}
\end{center}
\caption{
Spherically averaged density profiles of stars (solid line) and dark matter (dashed line) of the dwarf
galaxy in the initial configuration.
}
\label{initialprofiles}
\end{figure}

In this work we use a single model of a dwarf galaxy constructed numerically using the method of Widrow \& Dubinski (2005).
The dwarf is composed of an exponential stellar disk embedded in a cuspy,
cosmologically motivated Navarro et al. (1997, hereafter NFW) dark
matter halo. The halo had a virial mass
of $M_{\rm h} =10^9$ M$_{\odot}$ and a concentration parameter $c=20$.
The disk mass, $m_{\rm d}$, was specified as a fraction of 0.02 of the halo mass.
The disk radial scale length was $R_{\rm d} =
0.41$ kpc (Mo, Mao \& White 1998) and the disk thickness was determined by the
thickness parameter $z_{\rm d}/R_{\rm d} = 0.2$, where $z_{\rm d}$ denotes the
vertical scale height of the disk.
The dwarf galaxy model contained $N_{\rm h} = 10^6$ dark matter particles and $N_{\rm d} = 1.2 \times 10^6$ disk
particles. The gravitational softening was chosen to be $\epsilon_{\rm h}=60$~pc
and $\epsilon_{\rm d}=15$~pc for the two components,
respectively.
The initial density profiles of stars and dark matter in the dwarf measured in spherical shells are shown in
Figure~\ref{initialprofiles}.

The Milky Way model was
constructed as a live realization of the MWb model of Widrow \&
Dubinski (2005) with an exponential stellar
disk of mass $M_{\rm D}=3.53 \times 10^{10}$ M$_{\odot}$, a bulge
of mass $M_{\rm B}=1.18 \times 10^{10}$ M$_{\odot}$, and an NFW dark matter halo
of mass $M_{\rm H}=7.35 \times 10^{11}$ M$_{\odot}$.
Our model of the Milky Way had $N_{\rm D}=10^{6}$ particles
in the disk, $N_{\rm B}=5\times10^{5}$ in the bulge, and $N_{\rm H}=2\times10^{6}$
in the dark matter halo. The adopted gravitational
softenings were $\epsilon_{\rm D}=50$~pc, $\epsilon_{\rm B}=50$~pc, and
$\epsilon_{\rm H}=2$~kpc, respectively.

The dwarf galaxy model was placed on seven different orbits O1-O7 (see Table~\ref{parameters})
around the primary galaxy. The orientation of the internal angular momentum
of the dwarf with respect to the orbital angular momentum was mildly prograde
and equal to $i=45^{\circ}$. In all simulations, the dwarf was
initially placed at the apocentre of the orbit and the evolution was followed for $10$~Gyr using the
multistepping, parallel, tree $N$-body code PKDGRAV (Stadel 2001).

The orbits were chosen so as to probe a variety of sizes and eccentricities and their effect on the tidal tails
formed during the evolution.
The orbital apocentres, $r_{\rm apo}$, and pericentres, $r_{\rm peri}$, are
listed in the second and third column of Table~\ref{parameters}. In the first three orbits O1-O3 only the
size was varied while keeping the ratio $r_{\rm apo}/r_{\rm peri}$ (fourth column of Table~\ref{parameters})
constant and equal to 5, a value characteristic of subhaloes in cosmological environments of the size of the Milky Way
halo (Diemand et al. 2007). The sizes of orbits O1-O3 bracket those typically found for subhaloes in the Local Group
that survived until the present time (Klimentowski et al. 2010). In particular, the tight orbit O2 is similar to
the best estimates of the orbit of the Sagittarius dwarf (Law et al. 2005; {\L}okas et al. 2010) while the extended
orbit O3 is close to the one recently estimated for the Large Magellanic Cloud (Besla et al. 2007). The intermediate
orbit O1 can be considered as a typical orbit of a subhalo falling in at redshift $z = 1-2$. Orbits O4-O7 vary in
size and eccentricity so that both smaller and larger eccentricities are covered.

\begin{figure}
\begin{center}
    \leavevmode
    \epsfxsize=6.3cm
    \epsfbox[0 15 259 840]{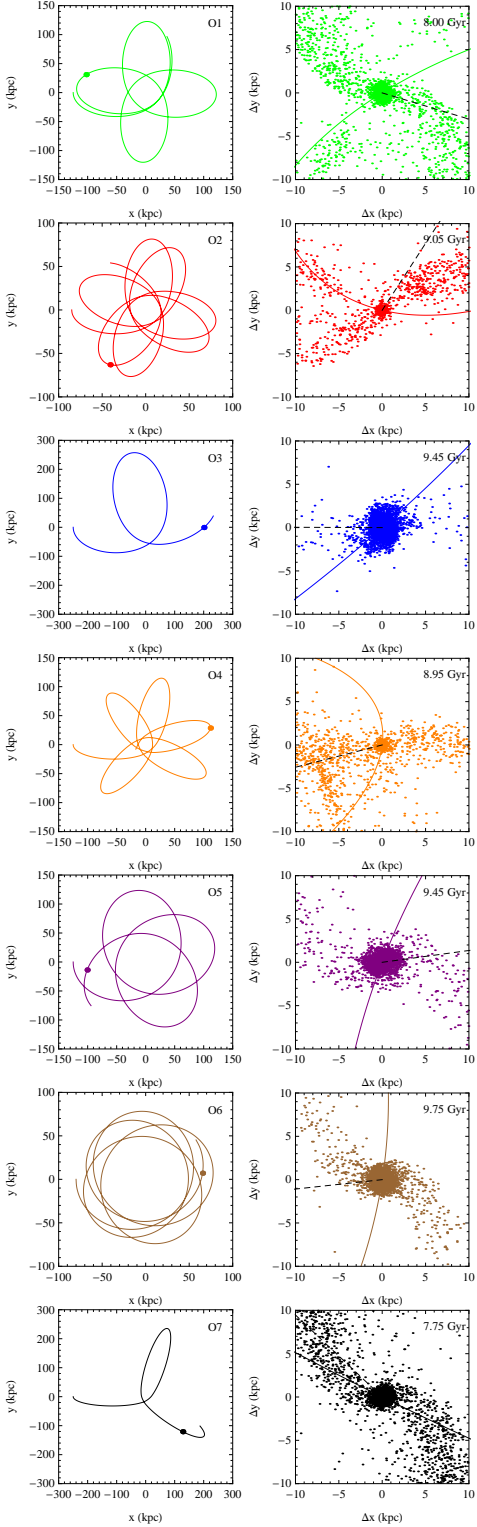}
\end{center}
\caption{Left-hand panels: orbits O1-O7 of the dwarf galaxy around the Milky Way projected onto the initial orbital plane.
Right-hand panels: magnified views of the stellar component of the dwarf galaxy and its tidal tails for orbits O1-O7.
The plots of the right column show 1 percent of stars at the time of the last occurrence of the maximum density
of the tails, given in the
upper right corner of each panel and marked with a dot in the corresponding panel of the left column. The coloured
solid line in all panels shows the orbit and the black dashed line in the right-column plots indicates
the direction toward the centre of the Milky Way.}
\label{orbitstails}
\end{figure}

The orbital times corresponding to all orbits are listed in the fifth column of Table~\ref{parameters}.
In the sixth column we provide the times of the last apocentre and in the seventh the number of pericentre passages
the dwarf experienced during the 10 Gyr of evolution. The last column gives the colour with which the results for
a given orbit will be shown throughout the paper. The left-column panels of Figure~\ref{orbitstails} show the
seven orbits considered in this work projected onto the initial orbital plane. Note that the orbits are not exactly
planar because the potential of the Milky Way, in which the dwarfs evolve, is not exactly spherical. The departures
from the sphericity, introduced by the presence of the Milky Way disk inclined by $45^\circ$ to the initial orbital
planes of the dwarfs makes the orbits deviate from these initial planes. This is especially the case for orbit O2 which
has a rather small pericentre and orbital time so that at the end of the evolution the departures are quite pronounced.

\section{The break radii}

The detailed picture of the evolution of a dwarf galaxy on an orbit around the Milky Way has been presented in
a few earlier works (e.g., Klimentowski et al. 2009b; Kazantzidis et al. 2011; {\L}okas et al. 2011) which focused
on the properties of the dwarf's main body. It has been demonstrated that as the dwarf
proceeds on its orbit, its stellar component transforms morphologically from a disk to a bar and then a spheroid while
the ordered stellar motions (the initial rotation) become more and more random. On tight enough orbits and after
long enough times a spheroidal galaxy is formed. This process is accompanied by strong mass loss, both in the stellar
and dark component. The material lost by the dwarf forms pronounced tidal tails on both sides of the dwarf: the leading
tail, traveling faster that the dwarf, and the trailing tail which moves slower. The tails do not follow exactly the
dwarf's orbit but rather move on their own orbits of different energy around the Milky Way.

In this section we look at the scales of transition between the main body of the dwarf and its tidal tails. This
transition is best quantified by measuring density profiles of the dwarf galaxy.
The density profile, both in stars and dark matter, rather than
steepening all the way to infinity, as is characteristic of systems unaffected by tides, steepens only out to
a certain scale and then flattens.
As will be discussed below, at maximum steepness the slope of the stellar density profile is close to $R^{-6}$
at all times, while at the outer radii it flattens to about $R^{-2}$. Let us therefore define the scale of transition
as a radius where the slope becomes less and less negative and crosses the value of $R^{-4}$.
In the following we will refer to this scale length of transition as the break radius.

In addition, the density profiles show a characteristic variation in time. Soon after
the pericentre passage the steepening occurs at the smallest radius and the flattened profile at larger
radii has a higher level of density. This happens when most of the material
is ejected from the dwarf by tidal forces which are strongest at the pericentre. Later on, however, the dwarf does not
stay truncated at the same scale but expands so that as the dwarf travels from the peri- to the apocentre
the break radius occurs at a larger and larger distance from the centre of the dwarf until, after the next
pericentre, a new strong steepening occurs and new material is fed into the tails. The increase of the break radius
is accompanied by decrease in the density of material outside, in the tidal tails.

\begin{figure}
\begin{center}
    \leavevmode
    \epsfxsize=6.6cm
    \epsfbox[0 15 245 812]{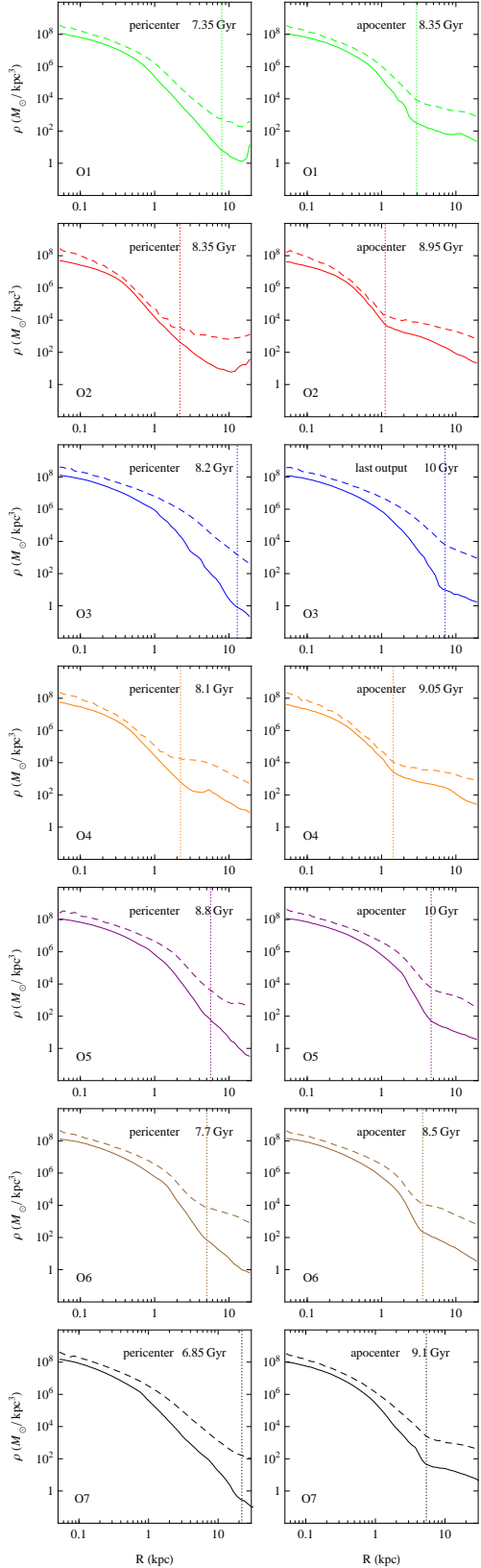}
\end{center}
\caption{Density profiles of the stellar component (solid lines) and dark matter (dashed lines) of dwarf galaxies
at the last apocentre of the orbit (right panels) and the preceding pericentre (left panels). The times of these
events are given in the upper right corner of each panel.
Vertical dotted lines indicate break radii.
}
\label{profiles}
\end{figure}

Figure~\ref{profiles} shows the density profiles of stars and dark matter (solid and dashed lines respectively)
measured in spherical shells around the dwarf galaxy centre. In the right-column panels we plot the density profiles
at the last apocentre on a given orbit (except for O3 where the last output is used instead because there are too
few apocentres), while in the left column we show density profiles at the preceding pericentre. The shapes of the density
profiles at these two instances are clearly different. At apocentres a clear break radius and the transition to the
tidal tails is seen. At pericentres no such clear transition is visible, there is rather a smooth transition from
the steeper to the shallower profile. The reason for this behaviour is that the material ejected at the previous
pericentre has already travelled away from the dwarf and a new ejection has not yet taken place.

\begin{figure}
\begin{center}
    \leavevmode
    \epsfxsize=8.2cm
    \epsfbox[0 0 233 507]{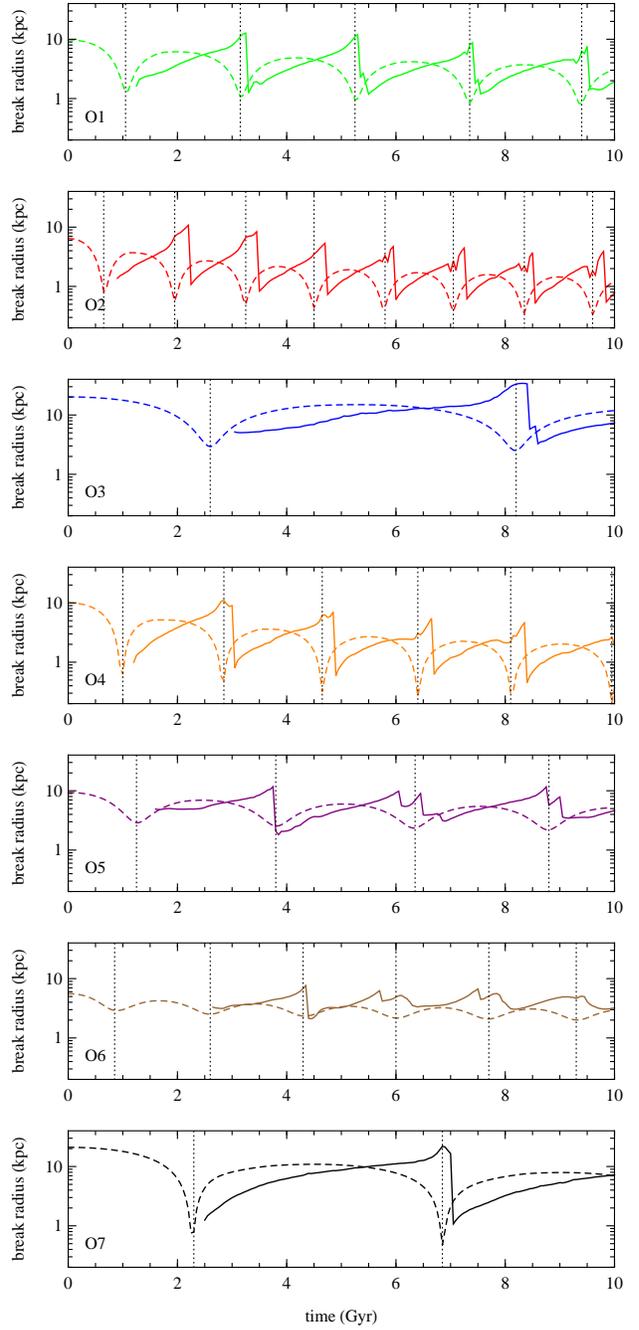}
\end{center}
\caption{Solid lines: the break radius of the stellar density profile as a function of time for different orbits,
measured directly from the simulations as the radius where the slope of the stellar distribution flattens.
Dashed lines: the tidal radius as predicted by formula (\ref{read}) with $\alpha=0$.
Vertical dotted lines indicate pericentre passages.}
\label{breakradius}
\end{figure}

We illustrate this behaviour further by plotting in Figure~\ref{breakradius} the break radii for dwarfs on all
orbits as a function of time (solid lines).
%From now on we will adopt as break radii the distances where the {\em stellar\/}
%density profile is steepest. In order to find this characteristic radius we calculate the slope of the stellar
%density profile by fitting a power law to each few adjacent density measurements and look for deep minima of the
%slope as a function of radius. Unfortunately, there is often more than one minimum of the slope due to non-sphericity
%of the stellar component in most orbits, especially at the early stages of evolution. A minimum may also occur
%further out in the tails. In order to distinguish between these situations we check if the density slope increases
%above $-3$ between minima which signifies the real transition to the tidal tails rather than the variation of the slope
%due to non-sphericity. The break radii calculated in this way for dwarfs on different orbits
%are shown in Figure~\ref{breakradius} as solid lines. One may argue that a break radius should rather be defined
%as a radius of transition between the steep density profile in the dwarf and the shallower one in the tails.
%This transition however happens on a range of radii and it would be difficult to say precisely where it occurs,
%while the minimum of the slope is mathematically well defined and can be determined accurately even from noisy
%numerical data.
The evolution of the break radii shows a characteristic pattern: as the dwarf galaxy expands after the pericentre
passage the break radius increases until well after the apocentre. As the
dwarf approaches the next pericentre the stellar distribution is stretched and the break radius increases
suddenly which manifests itself in peaks of break radii right after the pericentre in Figure~\ref{breakradius}.
The peaks are most pronounced and occur after all pericentre passages on orbits O2 and O4 which have the smallest
pericentres. Soon after the pericentre the stellar density is trimmed again and the break radius drops to its next
smallest value.

The break radii discussed here must obviously be related to the radius commonly referred to as the tidal radius.
For a satellite moving on a circular orbit, the tidal radius may be approximated as a Jacobi or Roche radius
defined as the last closed zero-velocity surface surrounding the satellite (Binney \& Tremaine 2008, sec. 8.3).
However, in the general case, also applicable to most dwarf galaxies orbiting the Milky Way, the orbits are
not circular and there is no reference frame in which the potential experienced by the stars in the dwarf
is stationary. In these cases no analog
of the Jacobi radius exists. A different approach to deriving the tidal radius was proposed by King (1962) who
argued that the limit can be set by finding a radius, on the line joining the host and the satellite, where
the acceleration of a star of the satellite vanishes. Read et al. (2006) generalized this derivation to take
into account the type of orbit of the star within the satellite. Their formula (18) can be rewritten as
\begin{equation}    \label{read}
	R_{\rm t} = r \left( \frac{M_{\rm s}}{M_{\rm g}} \frac{r}{\Lambda} \right)^{1/3}
	\left[ \frac{1}{\sqrt{1 + 2(r/\Lambda) + \alpha^2} + \alpha} \right]^{2/3}
\end{equation}
where $R$ is the distance measured from the dwarf's centre, $r$ is the distance between the centre of the Milky Way
and the centre of the dwarf, $M_{\rm s}$ and $M_{\rm g}$ are the masses of the satellite and the host galaxy
and $\alpha$ describes
the type of the star's orbit within the satellite, so that $\alpha = 1$, 0 and $-1$ respectively for prograde,
radial and retrograde orbits. The formula is strictly valid only for point-mass potentials of the satellite and
the host so that the orbit is Keplerian and $\Lambda = 2 r_{\rm apo} r_{\rm peri}/(r_{\rm apo} + r_{\rm peri})$.
Note that after setting $\alpha = 0$ and $r = r_{\rm peri}$ formula (\ref{read}) reduces to the well-known
King (1962) prescription for the tidal radius
\begin{equation}    \label{king}
	R_{\rm t} = r_{\rm peri} \left[ \frac{M_{\rm s}}{M_{\rm g} (3+e)} \right]^{1/3}
\end{equation}
where $e = (r_{\rm apo} - r_{\rm peri})/(r_{\rm apo} + r_{\rm peri})$.

King (1962) argued that once the satellite is trimmed down to this smallest tidal radius at the pericentre it will
preserve this shape also in the other parts of the orbit. As shown in Figure~\ref{breakradius}, this is not however
the case for our dwarfs: the break radii change with the position of the dwarf on the orbit. Instead of the
simple formula (\ref{king}) we thus used the more general expression (\ref{read}) adopting as the satellite mass the mass
contained within the tidal radius, $M_{\rm s}(R_{\rm t})$ and as the Milky Way mass the mass within the
current distance of the dwarf $M_{\rm g}(r)$. Given the orbits considered here the only important dependence
on the distance in $M_{\rm g}$ is due to the NFW-like distribution of the dark mass in the Milky Way halo thus
the disk and bulge are included as point masses. With these assumptions, equation (\ref{read}) provides an implicit
formula for $R_{\rm t}$ which can be solved numerically to arbitrary accuracy for any $\alpha$.
Note that replacing $r_{\rm peri}$ by $r$ in equation (\ref{king}) does not lead to the correct prescription
for the tidal radius and formula (\ref{read}) should be used instead.

The results of this semi-analytic procedure for the calculation of the tidal radius are shown as dashed lines in
Figure~\ref{breakradius}. Only the curves for the $\alpha=0$ are shown, since those agree best with the
values of the break radii determined from the simulations both in terms of the maximum and minimum values. We have
also considered $\alpha = 1$ and an intermediate value $\alpha = 0.5$ (see below) but they lead to systematically
lower predictions. The break radii of the simulated dwarfs are
always delayed with respect to the instantaneous tidal radii predicted by the formula, probably because of the
time the satellite needs to respond dynamically to the tidal forces exerted by the host galaxy at earlier times.
Thus the minimum of
the measured break radius occurs some time after the pericentre and then the break radius monotonically increases
until it reaches a plateau at the level close to the maximum tidal radius predicted by formula (\ref{read}) at
apocentres.
After that the transition to the tidal tails in the outer density profile moves to even larger radii leading
to even higher values of the break radii at and immediately after the pericentre while the predicted values
already reach another minimum.

Thus the break radii of the simulated dwarfs are shifted in time with respect to the predicted tidal ones by
a period the dwarf needs to dynamically respond to the tidal forces from the host. The global trend of break
radii decreasing over large time scales (of at least one orbital time), due to the mass loss of the dwarf is however
seen in both predictions and direct measurements. Both also
display the clear dependence on the eccentricity of the orbit: the difference between the maximum and minimum
break radius over one full orbit is significantly larger for eccentric orbits than more circular ones. In
particular, the almost circular orbit O6 shows very little variation in break radii over orbital time.

It is perhaps surprising that a simple formula (\ref{read}) reproduces reasonably well the general behaviour of
the break radii. We have verified that the prescriptions for the tidal radius taking into account
the dependence of slope of the host galaxy's mass distribution on radius do not improve the agreement.
As already mentioned, we also find
the agreement to be best for $\alpha=0$ which corresponds to the radial orbits of the stars in the dwarf galaxy.
Although the stellar motions are initially mildly prograde with respect to the orbits (the initial disks are inclined
by 45$^\circ$ to the orbital plane), so that $\alpha = 1$ or an intermediate value $\alpha = 0.5$ would seem more
appropriate, in most cases after the first pericentre passage a bar is formed which is dominated
by radial motions of the stars. This may explain why $\alpha=0$ in formula (\ref{read}) reproduces the simulation
results better than other $\alpha$ values.

\begin{figure}
\begin{center}
    \leavevmode
    \epsfxsize=8.43cm
    \epsfbox[0 0 233 494]{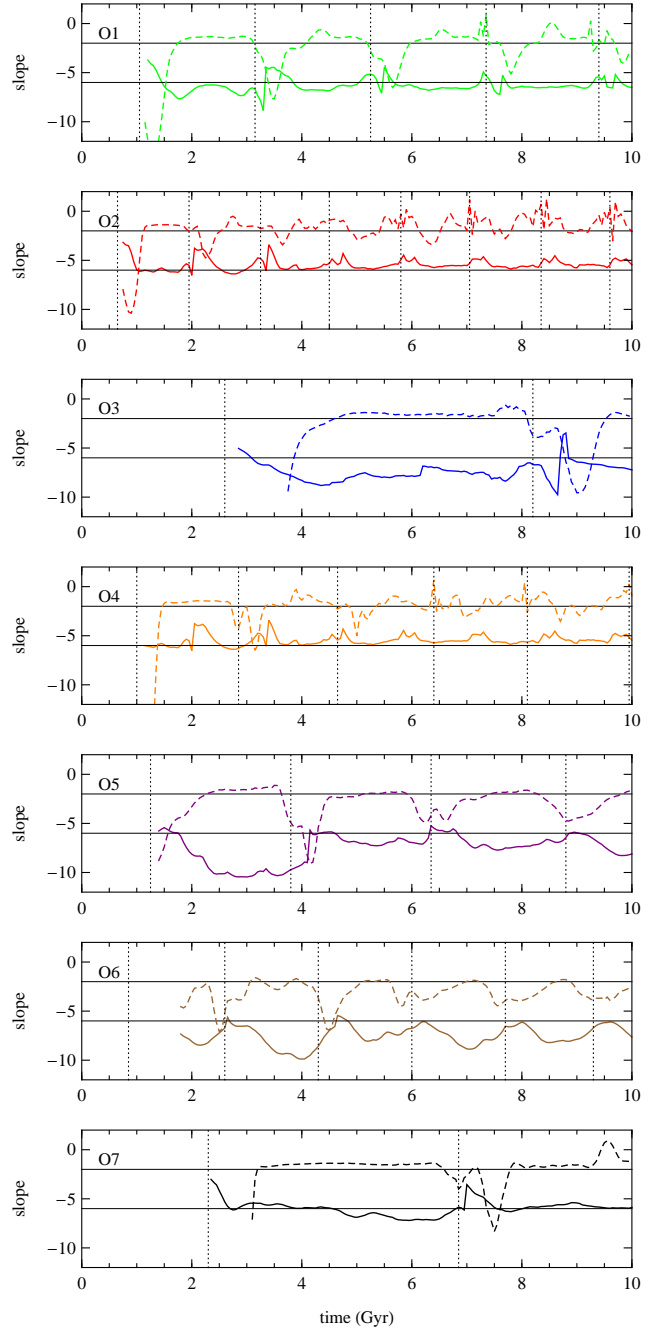}
\end{center}
\caption{Solid lines: the minimum slope of the stellar density profile near the break radius as a function of time
for different orbits. Dashed lines: the slope of the stellar density profile in the tails measured near $R_{\rm tail}$.
Two thin horizontal lines mark the slopes of $-2$ and $-6$.
Vertical dotted lines indicate pericentre passages.}
\label{slope}
\end{figure}

In Figure~\ref{slope} we plot with solid lines the actual minimum slope of the stellar density profile which is
found at a radius somewhat lower than the
break radius.
Interestingly, this slope turns out to be remarkably constant in time (between pericentres)
and close of $-6$, a value marked with one of the thin horizontal lines in each panel. This slope is particularly
well preserved for orbits O2 and O4 where the stellar component becomes almost perfectly spherical early on
during the evolution, but is also quite constant for the most eccentric orbit O7 and the typical one O1.
This finding thus provides justification to analytic formulae proposed to describe the stellar density
distribution in dwarf galaxies and globular clusters with power-law behaviour at large radii. In particular,
the slope we find is close to the one in a commonly used formula by Plummer (1915) where the
density varies as $R^{-5}$ far away from the centre of an object. In orbits O3, O5 and O6 the slope is
however significantly lower for most of the evolution, and thus the stellar density profiles would be better
fitted by formulae where density decreases exponentially with radius like the S\'ersic (1968) law or
the profile more recently proposed by Kazantzidis et al. (2004). Note also, that the formula proposed
by King (1962), where the dwarf is tidally truncated to zero density at some radius, does not provide a good
description of the density distribution of our simulated dwarfs.

\section{The density and kinematics of the tails}

In the following analysis of the properties of the tidal tails we will assume that they can be reliably
measured at
$R > R_{\rm tail}$
where $R_{\rm tail}$ is the maximum value of the break radius occurring near
the second pericentre.
%, where its values reach a first maximum or flatten as a function of time. In determining
%these values we exclude the times near pericentres when, as discussed above, the stellar density profiles
%are extended and the break radii are not well defined.
The values of $R_{\rm tail}$ found in this way
are listed in the eighth column of Table~\ref{parameters}. The adopted $R_{\rm tail}$ will be different for
each orbit, but kept constant in time. Since the maximum break radius decreases in subsequent pericentres, this
means that $R_{\rm tail}$ is always significantly larger than the instantaneous break radius.

One of the basic properties that can affect the modelling of dSph galaxies and also possibilities of detection
of tidal tails around them is the density of stars in the tails. We measured this density as a function of time
by counting stars within a shell of radii $(R_{\rm tail}, R_{\rm tail} + \Delta R)$ where we adopt
$\Delta R=1$ kpc except for the case O3 where
we take a thicker shell of $\Delta R=6$ kpc because the tails are very weak in this case.
These choices guarantee that in all outputs (after the first pericentre) we have at least 10 stars in the shell
and the measurements are meaningful. Since the volume of the shell is therefore different for each orbit, to
facilitate comparison between different orbits we
normalize the measurements of density to the lowest value found near the second pericentre, where the tails in
all cases are already well formed (note that the second pericentre is also the last for orbits O3 and O7).

\begin{figure}
\begin{center}
    \leavevmode
    \epsfxsize=8.2cm
    \epsfbox[0 0 232 498]{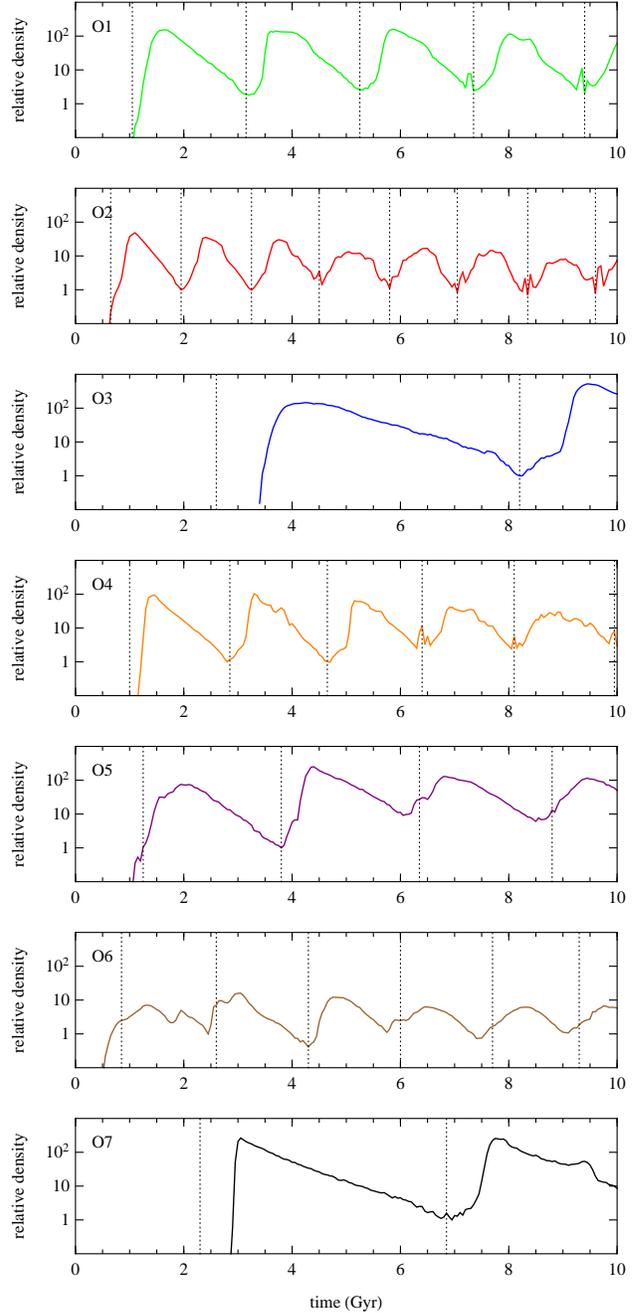}
\end{center}
\caption{The relative density of tidal tails at $R_{\rm tail}$ as a function of time for different orbits. The values
were normalized to the minimum density near the second pericentre of each orbit. Vertical dotted lines indicate
pericentre passages.}
\label{density}
\end{figure}

The relative density calculated in this way is shown in Figure~\ref{density}. The evolution of the density
shows a characteristic pattern which is preserved at all times and for all orbits: it has a minimum at or
just before pericentre, and a maximum after the pericentre passage but before the apocentre is reached. The
density at the maximum is between 10 and a few hundred times larger than at the minimum and this factor depends strongly
on the orbit. It is smallest for the most circular orbit O6 and among the
largest for the most eccentric one O7. In addition,
the maximum occurs at a smaller fraction of the orbital time after the pericentre passage for more eccentric
orbits like O7; in this case the density in the tails grows steeply in time and then decreases more slowly
towards the next pericentre. This behaviour is obviously due to stronger mass ejection during pericentre passage
on eccentric orbits where the dwarf is tidally shocked. Interestingly, the density of the tails changes
significantly (by a factor of 10) over orbital time even for the most circular orbit O6. Here however, the
phases of increasing and decreasing density last for a comparable time.

\begin{figure}
\begin{center}
    \leavevmode
    \epsfxsize=8.32cm
    \epsfbox[0 0 232 492]{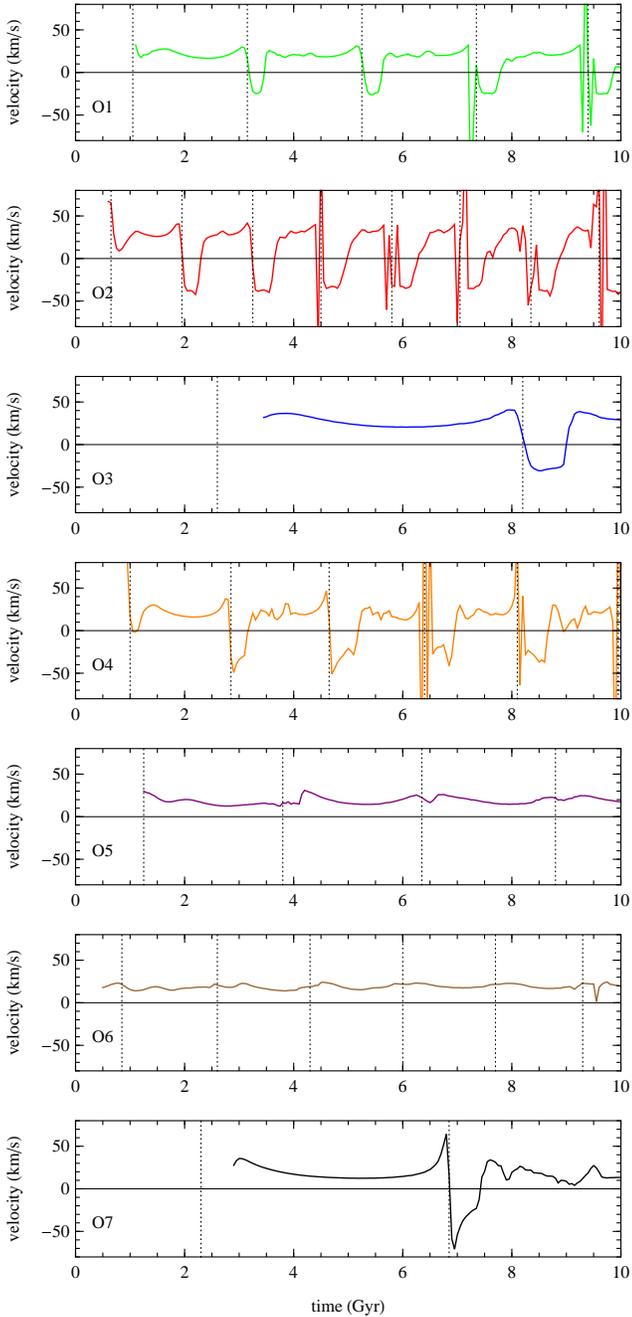}
\end{center}
\caption{The radial velocity of tidal tails with respect to the dwarf at $R_{\rm tail}$
as a function of time for different orbits. Vertical dotted lines indicate pericentre passages.}
\label{velocity}
\end{figure}

It is also interesting to measure the slope of the stellar density distribution near $R_{\rm tail}$. The values
of this slope are plotted in Figure~\ref{slope} as dashed lines. It turns out
to be remarkably close to $-2$ (marked as one of the thin horizontal lines in each panel of the Figure)
and rather constant as a function of time (except for pericentre passages), but
also uniform between orbits. This value is a direct consequence of the density of the stars being constant
along the tail. The tails are thus well approximated by cylinders of constant density. When such density
distributions are probed in spherical shells the resulting slope is $-2$ as we find by direct measurement.
The measurements are a little more noisy for orbits O4 and O6 where debris from multiple pericentre passages
overlap.

The intuitive picture of the tidal tails with the leading tail preceding the dwarf on the orbit and the trailing
tail travelling behind would suggest that viewed from the centre of the dwarf galaxy, the stars in the tails
should always recede. It is worthwhile to verify this statement by directly measuring the radial velocities
of the tails within shells $(R_{\rm tail}, R_{\rm tail} + \Delta R)$ as before. We define positive velocities as
those receding from the dwarf, and negative and those approaching the dwarf. The results of the measurements
(using the same selection of stars as for the determination of density) are shown in Figure~\ref{velocity}.

It turns out that the stars at $R_{\rm tail}$ indeed move away from the dwarf for most of the time on all orbits,
but they do so {\em at all times\/} only for the least eccentric orbits O5 and O6. For other orbits there are periods
right after the pericentre passage when the stars in the tails stop with respect to the dwarf or even approach
it (negative velocities). These dips of velocity are most pronounced for the most eccentric orbits O7 (after the second
pericentre) and O4 (after the second and third pericentre), but are also well visible for the less eccentric
orbits O1-O3.
Note that the strong oscillations of velocity at later pericentres in orbits O1, O2 and O4 are due to
the contribution from the debris lost much earlier. The stars in these debris would have similar properties
as the stars lost recently and thus would be undistinguishable for observers. Therefore the velocity pattern discussed
above may not be observable if there are many wraps of debris lost by a given dwarf.

The interpretation of these negative velocities becomes clear once we consider the dynamics of the tails
in the vicinity of the dwarf. The material in the leading tail moves faster than the dwarf and thus
it reaches its pericentre sooner than the dwarf's centre. The opposite is true for the trailing tail: it moves
slower on its orbit and reaches the pericentre later. For very eccentric orbits, at the time when
the dwarf has just passed the pericentre, the leading
tail is already after the pericentre passage and its velocity is lower than the dwarf's while the trailing
tail is just reaching the pericentre and its velocity is larger than the dwarf's. Thus the stars near $R_{\rm tail}$
in both tails approach the dwarf.

\section{The orientation of the tails}

Another essential factor affecting the modelling of dwarf galaxies and the possibility of studying their
tidally stripped debris is the orientation of the tails with respect to an observer located near the centre
of the host galaxy. Here we assume that the observer is placed exactly at the centre of the Milky Way since
the Sun's position's offset with respect to it is small in comparison with the distances of most dwarf galaxies
orbiting the Galaxy.

\begin{figure}
\begin{center}
    \leavevmode
    \epsfxsize=8.5cm
    \epsfbox[0 0 232 503]{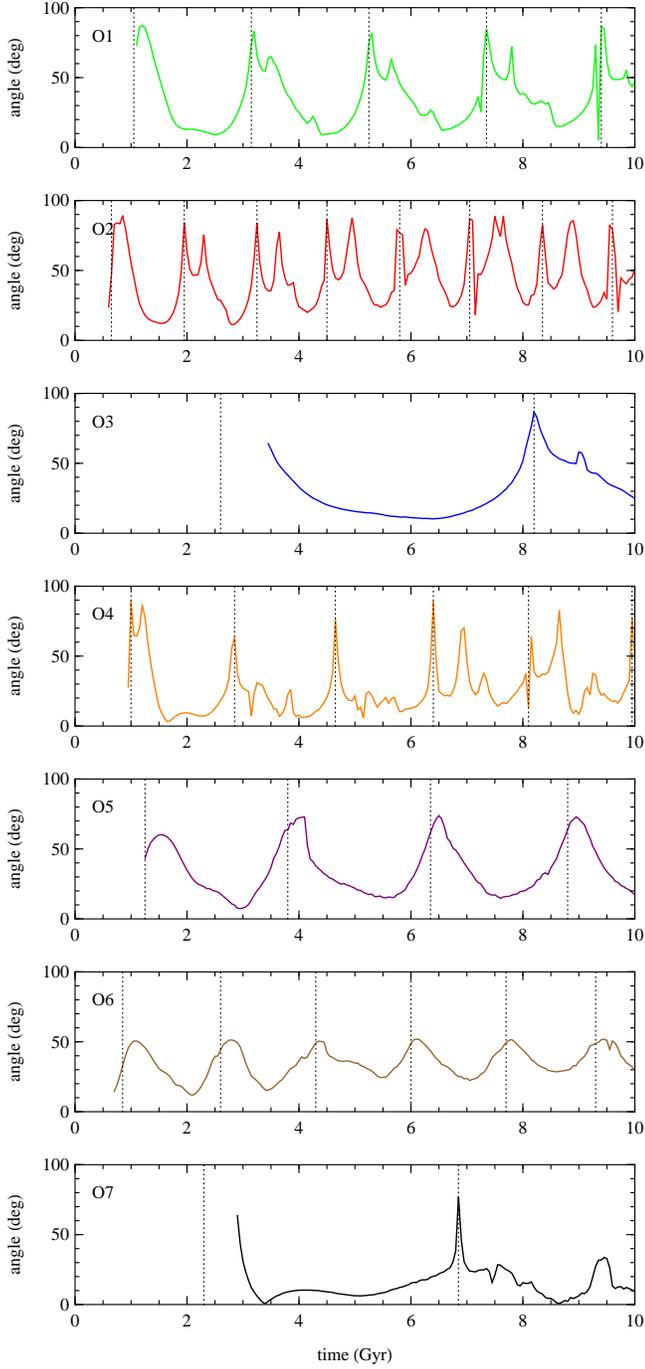}
\end{center}
\caption{The angle between tidal tails and the direction toward the centre of the Milky Way at $R_{\rm tail}$
as a function of time for different orbits. Vertical dotted lines indicate pericentre passages.}
\label{angle}
\end{figure}

\begin{figure}
\begin{center}
    \leavevmode
    \epsfxsize=4.95cm
    \epsfbox[0 12 170 772]{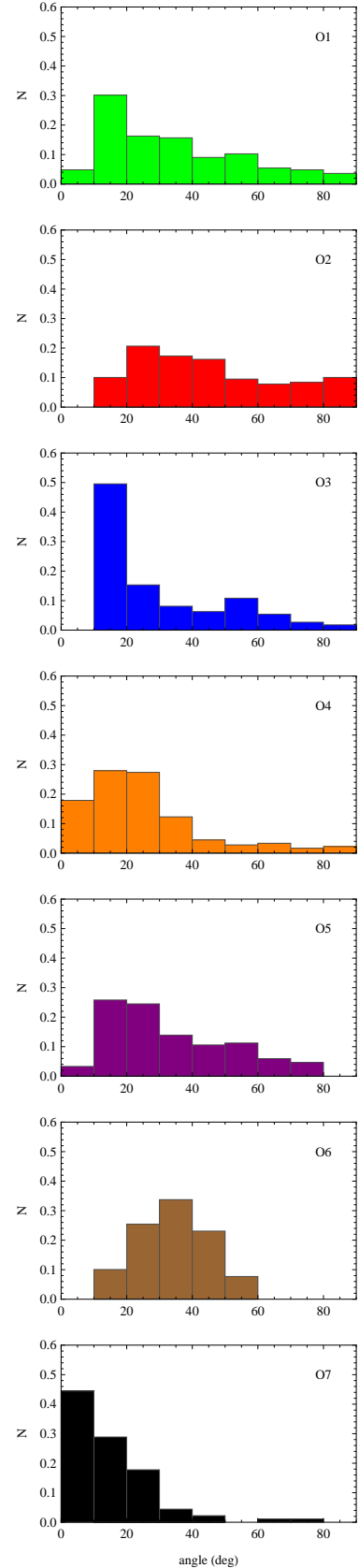}
\end{center}
\caption{The distribution of the angle between tidal tails and the direction toward the centre of the Milky Way
for different orbits. The histograms were constructed from outputs between the first and last pericentre and
normalized to unity.}
\label{histangle}
\end{figure}

The stellar component of the dwarf galaxy and its tidal tails in the immediate vicinity
(up to $R=10$ kpc) is shown (projected onto the initial orbital plane) for different orbits
in the right-column panels of Figure~\ref{orbitstails}.
The outputs pictured were chosen as those when the last maximum of the tails density occurs
(see Figure~\ref{density}) and the time values are given in each panel. The dashed black line in each panel
indicates the direction towards the Milky Way at this instant of evolution. The corresponding positions of the
dwarf on the orbit at these times are marked as coloured dots in the left-column panels
of Figure~\ref{orbitstails} showing the orbits.

In Figure~\ref{angle} we plot the angle between the tails and the direction to the centre of the Milky Way
as a function of time. The line of the tails was determined by selecting as before stars in
shells $(R_{\rm tail}, R_{\rm tail} + \Delta R)$ and finding the major axis of their distribution from the
tensor of inertia. In calculating the angle we always consider the tail closer to the direction toward
the Milky Way, so that the values of the angle are always between $0^\circ$ and $90^\circ$.

The evolution of the orientation of the tails in time shows an interesting regularity: for all orbits the highest
values of the angle are found near the pericentre. These highest values are close to 90 degrees for eccentric
orbits which means that the tails can be perpendicular to our line of sight only for a very short period of time.
For the most circular orbits O6 and O5, these high angles are never reached. In these two cases the tails can
be observed at most at about $50^\circ$ and $70^\circ$, respectively. For most of the time, however, the tails would be
observed at much smaller angles. In particular, at apocentres where the dwarfs spend most of the time, the angles
are very close to zero (except for the most circular orbit O6 where they are a little larger),
i.e. the tails are oriented almost exactly along the line of sight.

The picture is somewhat less clear for orbits O4 and O7 after the second pericentre. The noisy measurements at
these times are due to the fact that for these very eccentric orbits the material lost at earlier pericentre
passages and the tails formed after the latest pericentre are close in terms of radial direction. On the other
hand, the noisy measurement at later stages of orbit O2 are due to the tightness of the orbit (the newly formed
tails are very short and the dwarf passes the previously lost debris a few times).

The probability of observing the tails at a given angle can be illustrated by plotting histograms of the number of
occurrences of a given orientation as a function of the measured angle, as we do in Figure~\ref{histangle}.
For each orbit we included measurements for the available number of full orbits between the first and the last
pericentre (or a single orbit slightly shifted forward in time for orbits O3 and O7 where no measurements are
available immediately after the first pericentre). The histograms were then normalized to unity. It is clear that
the most probable angle of tidal tail orientation is below or very close to $20^\circ$ for all orbits,
except the most circular
orbit O6, where it is below 40 degrees. In addition, all orbits have a very weakly populated tail of the
distribution at large angles (orbit O6 has no such tail at all).

As in the case of the velocity of the tails, this behaviour can be understood based on the dynamics of the tails
as they travel along the orbit. As already discussed by Klimentowski et al. (2009a), after the pericentre passage
most of the stars newly stripped from the dwarf remain close to the dwarf's main body, but travel on their own orbits.
As they approach the apocentre of the orbit, the leading tail slows down earlier than the trailing tail and the
orientation of the tails becomes more aligned with the direction towards the centre of the Milky Way. The opposite
is true on the way from the apo- to the pericentre of the orbit: the tails are stretched along the orbit and their
orientation becomes more perpendicular to the direction towards the Milky Way.

\section{Discussion}

We have studied the properties of tidal tails forming around dwarf galaxies orbiting a Milky Way-like host. By
measuring the density profiles of the stellar components of the dwarfs we have
shown that the dwarf galaxies are not truncated at tidal radii imposed at pericentres but rather expand on their way
from the peri- to the apocentre. The transition to the tails occurs later than predicted by the formulae for the tidal
radius.
Our conclusions concerning the break radii and their relation to the tidal radii agree with the recent findings
of Webb et al. (2013) for globular clusters.

The process of mass loss is not instantaneous and does not occur
only at the pericentre, even for most eccentric orbits. Instead, the lost material is fed into the tails
gradually and the maximum density of the tails is reached a substantial fraction of orbital time after the pericentre
passage. The shape of the stellar density profile is similar at all orbital phases and all orbits: the outer profile
of the dwarf's main body follows an $R^{-6}$ law for most of the time, while the density distribution in the tails is
well approximated by $R^{-2}$, a slope consistent with the debris having constant density along the tail.

\begin{figure}
\begin{center}
    \leavevmode
    \epsfxsize=7cm
    \epsfbox[0 10 160 320]{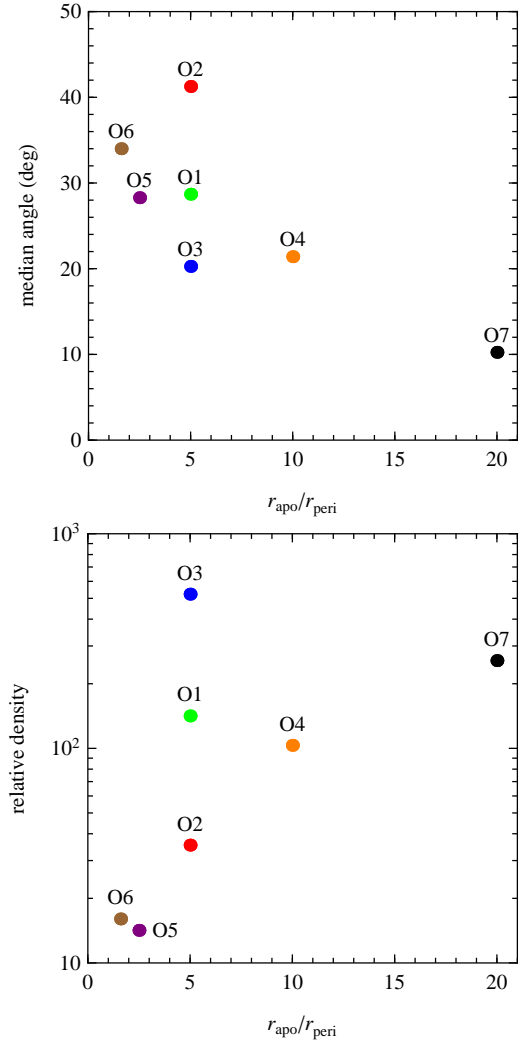}
\end{center}
\caption{The median angle between the dwarf galaxy's tidal tails and the direction toward the Milky Way's centre
(upper panel) and the maximum density of the tails near the third apocentre with respect to the minimum near the
second pericentre (lower panel) as a function of the orbit's eccentricity $r_{\rm apo}/r_{\rm peri}$. }
\label{angledens}
\end{figure}

Figure~\ref{angledens} summarizes our results in terms of the dependence of the two crucial quantities on the
eccentricity of the dwarf's orbit around the host. In the upper panel of the Figure we show the median of the angle
between the tidal tails of the dwarf and the direction to the centre of the Milky Way, measured from the distributions
shown in Figure~\ref{histangle}. The coloured dots, marked with the orbit symbol, are plotted as a function of
$r_{\rm apo}/r_{\rm peri}$. We can see the clear trend of decreasing median angles indicated by points
corresponding to orbits O6, O5, O1, O4 and O7 as we go towards the increasing eccentricity. The range encompassed by
the three orbits of the same eccentricity, $r_{\rm apo}/r_{\rm peri} =5$, namely, O2, O1 and O3 may be considered as
a measure of the scatter for this angle-eccentricity relation due to different sizes of the orbits. The fact that the
median angle is significantly lower for O3 and higher for O2 in comparison to O1 is related to the tightness
of the orbits.
For the tight orbit O2 the stripping is much more effective and there is much more stripped material in the vicinity of
the dwarf at any time than for the intermediate orbit O1 and especially so for the most extended orbit O3. In the case
of O3 there are very few stars lost, the tails are particularly clear-shaped and the material lost recently never
overlaps with the stars lost much earlier. Note also that the median angle never exceeds 42 degrees, even for the
most circular orbit O6.

In the lower panel of Figure~\ref{angledens} we plot in a similar way the relative density of the tails at the maximum
occurring near the third apocentre of each orbit measured with respect to the minimum reached near the second pericentre,
i.e. the pericentre used for normalization in Figure~\ref{density}. The only exception is orbit O6
which shows a little anomalous behaviour at the second pericentre (much lower density) compared to the following
pericentres. For this case we use the maximum density near the fourth apocentre with respect to the minimum at the
third pericentre. The relation between the relative density and eccentricity of the orbit is also visible, although
it is not as clear as the one for the angle. While the overall trend is to have increasing density with eccentricity,
the case of orbit O3 stands out somewhat. This is probably because the amount of stars lost on this orbit is the
smallest and the minimum at the second pericentre is lower than the next ones would be (had we evolved the dwarfs for
a period longer than 10 Gyr).

Our findings have immediate consequences for the studies of dSph galaxies in the Local Group. We predict that the
tidal tails around these galaxies may be extremely difficult to detect due to their orientation with respect to the
direction to the Milky Way. Given our proximity to the Galaxy's centre, this direction is approximately the same
as our line of sight if we take into account the considerable distances of most dSph galaxies.
We have shown that the tails are
typically inclined to this line of sight by relatively small angles, with median values always below 42 degrees.
The tails are oriented perpendicular to our line of sight only for rather short periods when the dwarfs are
near their orbital pericentres. Unfortunately, at these times the density of the tails is also at its lowest values
which makes the detection really improbable. The only exception from this rule may be the case of the Sagittarius dwarf
which indeed is near its pericentre at present but also close enough and on a tight enough
orbit to have a lot of stars stripped in the past so that its tidal tails could be detected over a large angular
range (Majewski et al. 2003).

Our results also emphasize the possible issues concerning the dynamical modelling of dSph galaxies. Such modelling
relies on the use of the samples of stellar velocities and can be trusted only in the case where most of the
velocities are indeed those of the stars still bound to the dwarf as only those are true tracers of the underlying
gravitational potential. As discussed thoroughly by Klimentowski et al. (2007), when the dwarf's tidal tails are
oriented along the line of sight the contamination of the kinematic samples can be particularly high leading to
overestimates of the mass and/or biases in the inferred density distributions and orbital properties. We have
shown that this will almost always be the case as dwarfs spend most of the time near apocentres of their orbits and
in addition this is also where the density of the tails is large. A particularly dangerous orbital stage from the
point of view of contamination is when the dwarf galaxy is on its way towards the apocentre, because this is when the
density of the tails reaches its maximum.

\section*{Acknowledgements}

This research was partially supported by the Polish National Science
Centre under grant NN203580940. GG acknowledges the summer student program of the Copernicus Center
in Warsaw.

\end{document}